\documentclass{article} 
\usepackage{iclr2026_conference,times}

\usepackage{enumitem}
\usepackage{algorithm}
\usepackage{algorithmic}
\usepackage{amsmath}
\usepackage{graphicx}
\usepackage{adjustbox,multirow}
\usepackage{arydshln}
\usepackage{float}

\usepackage{amsmath,amsfonts,bm}

\def\eqref#1{equation~\ref{#1}}

\def\1{\bm{1}}

\def\ri{{\textnormal{i}}}

\def\rs{{\textnormal{s}}}

\def\ru{{\textnormal{u}}}

\DeclareMathAlphabet{\mathsfit}{\encodingdefault}{\sfdefault}{m}{sl}
\SetMathAlphabet{\mathsfit}{bold}{\encodingdefault}{\sfdefault}{bx}{n}

\usepackage{hyperref}
\usepackage{url}
\usepackage{adjustbox,multirow}
\usepackage{booktabs}       

\title{SK\textsuperscript{2}Decompile: LLM-based Two-Phase Binary Decompilation from Skeleton to Skin}

\author{Hanzhuo Tan, Weihao Li, Xiaolong Tian, Siyi Wang, Jiaming Liu, Jing Li, Yuqun Zhang
}

\newcommand{\methodori}{SK\textsuperscript{2}Decompile}
\newcommand{\method}{\textit{SK\textsuperscript{2}Decompile}}

\newcommand{\struc}{Structure Recovery}
\newcommand{\ident}{Identifier Naming}

\iclrfinalcopy 
\begin{document}

\maketitle

\begin{abstract}
Large Language Models (LLMs) have emerged as a promising approach for binary decompilation. However, the existing LLM-based decompilers are still somewhat limited 
in effectively presenting a program's source-level structure with its original identifiers.
To mitigate this, we introduce \method{}, a novel two-phase approach to decompile from the skeleton (semantic structure) to the skin (identifier) of programs. Specifically, we first apply a \struc{} model to translate a program's binary code to an Intermediate Representation (IR) as deriving the program's ``skeleton'', i.e., preserving control flow and data structures while obfuscating all identifiers with generic placeholders. 
We also apply reinforcement learning to reward the model for producing program structures that adhere to the syntactic and semantic rules expected by compilers.
Second, we apply an \ident{} model to produce meaningful identifiers which reflect actual program semantics as deriving the program's ``skin''. We train the \ident{} model with a separate reinforcement learning objective that rewards the semantic similarity between its predictions and the reference code. 
Such a two-phase decompilation process facilitates advancing the correctness and readability of decompilation independently.
Our evaluations indicate that \method{} significantly outperforms the SOTA baselines, achieving 21.6\% average re-executability rate gain over GPT-5-mini on the HumanEval dataset and 29.4\% average R2I improvement over Idioms on the GitHub2025 benchmark.
\end{abstract}
\section{Introduction}
Decompilation refers to converting compiled binaries back to high-level source code and has been widely adopted in software security tasks like malware analysis and vulnerability discovery~\citep{decompilation1, decompilation2_rnn, decompilation3_dnn, decompilation4_gnn,decompilation4-coda}. Ideally, a decompiler ensures both functional correctness and code readability, which can hardly be realized in practice at the same time. For instance, traditional tools like Ghidra~\citep{ghidra} and IDA~\citep{idapro} excel at functional correctness but often produce obfuscated, hard-to-read code, while recent Large Language Model (LLM)-based approaches~\citep{btc,slade,nova,LLM4Decompile,self-decompile,idioms} generate more readable output but frequently fail to preserve the original program's functionality
~\citep{LLM4Decompile,decompilebench}.

Many research efforts imply the root cause of this trade-off as the intractable complexity of simultaneously inferring control-flow structures, data layouts, and identifiers in a single phase~\citep{dire,resym,dirty,binbench,name-recover1, name-recover2,type-recover1}. To mitigate this, we introduce \method{}, a novel LLM-based decompilation technique that decomposes the binary decompilation task into two phases. In particular, we first derive the program's skeleton, i.e., its core structure, including control flow and data structure~\citep{DragonBookControlFlow}. Then, we derive the program's skin, i.e., the meaningful type, variable, and function names reflecting the actual program semantics~\citep{dire}. Such a two-phase decompilation design allows for tackling the challenges of functionality and readability independently for aggregating their respective effectiveness rather than realizing a trade-off in between. In particular, we design a novel Intermediate Representation (IR) acting as the ``skeleton'' of the program. This IR essentially refers to the original source code with all identifiers (variable, function, and type names) replaced by generic placeholders~\citep{dobf-cite} for preserving structural and functional logic of a program, following the Information Bottleneck principle~\citep{infobot1,infobot2}. 
The decompilation process is then split into two sequential phases: \textit{\struc{}} where an LLM translates the compiled binary code to our structural IR and \textit{\ident{}} where a second LLM enriches the IR by predicting meaningful names reflecting actual program semantics for all placeholders. For \struc{}, we first train a sequence-to-sequence model~\citep{llmcompiler,transformers} and further tune it with reinforcement learning (RL)~\citep{gpt4}, where the compiler checks the syntax and semantics to provide the reward. A positive reward is generated only if the generated IR successfully compiles, with additional rewards reflecting the correctness of placeholder recovery. 
For \ident{}, we use a separate RL reward. To better capture human-centric readability, this model is not rewarded for exact name match but for the semantic similarity between its output and the reference code~\citep{qwen3embedding}. In this way, \method{} enhances functional correctness and semantic readability simultaneously for LLM-based decompilation.

Our evaluations show that \method{}
significantly outperforms prior SOTA models on four open-source benchmark suites. To our best knowledge, \method{} is the first to approach the average re-executability rate of $\sim$70\% on HumanEval~\citep{humaneval} and $\sim$60\% on MBPP~\citep{mbpp}. It also achieves 21.6\% average re-executability rate gain over GPT-5-mini~\citep{gpt5} on HumanEval and 29.4\% average R2I~\citep{r2i} improvement over Idioms~\citep{idioms} on the GitHub2025 benchmark~\citep{decompilebench}.  

The code has been released in anonymous GitHub page~\footnote{\url{https://github.com/albertan017/LLM4Decompile}}. 
Our main contributions are as follows. 

\begin{itemize}[leftmargin=*, topsep=0pt]
    \item \textbf{Two-phase Decompilation Framework.} We propose the first decompilation framework consisting of two phases: \struc{} for advancing the recovery of source-level program structures and \ident{} for advancing the recovery of meaningful identifiers reflecting actual program semantics. Each phase trains a model using reinforcement learning with specific rewards respectively. 

    \item \textbf{Intermediate Representation (IR).} We propose our IR as the obfuscated source code.
    This IR satisfies the Information Bottleneck principle by maximizing the compression of the semantics embodied in identifiers while preserving the semantics embodied in the structure of the program, and it is practically simple to generate.

    \item \textbf{Extensive Evaluations.} We perform extensive evaluations on \method{} and find out it achieves the optimal performance compared with the studied baselines. For instance, \method{} achieves 21.6\% average re-executability rate gain over GPT-5-mini on HumanEval and 29.4\% average R2I improvement over Idioms on the GitHub2025 benchmark.
\end{itemize}

\section{Background}
\subsection{Related Work}
Decompilation, i.e., the reconstruction of source code from binary executables, has long relied on control/data-flow analysis and pattern matching~\citep{decompilation1, decompilation2_rnn, decompilation3_dnn,decompilation4-coda}. 
Typically, conventional decompilers like IDA Pro~\citep{idapro} tend to recover a program's basic logic, with their generated pseudocode close to low-level assembly code, i.e., their outputs often lack readability and re-executability
~\citep{eval-decom1,eval-decom2}.

Motivated by the success of Large Language Models (LLMs) in code-related tasks~\citep{codestudyzheng,clap,binaryai,bianryanalysis,SALT4Decompile,codecompile}, recent research has focused on applying LLMs to refine the pseudocode generated by traditional decompilers~\citep{hu2024degpt,decgpt}. Note that as pseudocode is deterministic with the corresponding binary code, we use the terms interchangeably in this paper. 
Initial efforts, such as LLM4Decompile~\citep{LLM4Decompile}, demonstrated that LLMs could effectively learn to translate low-level pseudocode to high-level source code and inspire subsequent studies~\citep{self-decompile,ref-decompile}. Other research focuses on incorporating contextual information. For instance, Idioms~\citep{idioms} enriches the input by incorporating information from adjacent functions in the call graph and attempts to jointly recover user-defined type definitions with the decompiled code. Recently, D-LIFT~\citep{dlift} enhanced the training pipeline by incorporating reinforcement learning, guided by a novel reward function D-SCORE which provides a multi-faceted assessment of code based on accuracy and readability. Despite these advancements, the functional correctness of LLM-based techniques remains a significant challenge, with existing models failing on approximately half of the tasks in the HumanEval-Decompile benchmark~\citep{LLM4Decompile}.

\begin{figure*}[hbt]
  \centering
  \includegraphics[width=\textwidth]{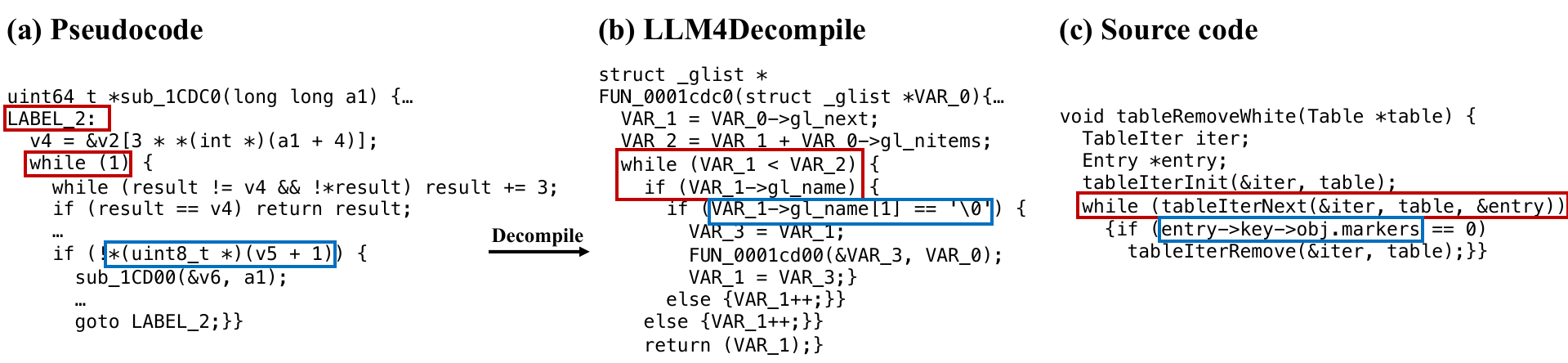}
  \caption{An example with its (a) pseudocode, (b) refinement by LLM4Decompile, and (c) source code. 
  \fcolorbox{red}{white}{red} marks the \texttt{while} loop in different forms, \fcolorbox{blue}{white}{blue} represents the \texttt{data access}.}.
  \label{fig:case}
\end{figure*}

\subsection{Motivating Example}
As shown in Figure~\ref{fig:case}, while LLM4Decompile, a widely-studied LLM-based decompiler, correctly interprets constructs like \verb|while(1)| and \verb|goto LABEL_2| found in the IDA pseudocode and successfully recovers them to a semantically equivalent and more readable \verb|while| loop.
However, the decompiler struggles with recovering the program’s data type structure, and its ability to assign meaningful identifier names reflecting actual program semantics remains limited. For instance, domain-specific types such as \verb|Table| and \verb|Entry| are erroneously mapped to a generic struct named \verb|_glist|. This fundamental limitation in the decompiler's understanding of the data organization leads to the failure in generating meaningful identifiers. Consequently, variables and functions are reduced to generic placeholders like \verb|VAR_1| and \verb|FUN_0001cdc0|, making it even more difficult to understand the original intent of the program. Such deficiencies motivate a two-phase decompilation process for recovering both program structure and meaningful identifiers respectively rather than realizing a trade-off in between, as illustrated in Section~\ref{approach}.

\section{\methodori{}}
\label{approach}
\subsection{Overview}
\begin{figure*}[hbt]
  \centering
  \includegraphics[width=1.0\textwidth]{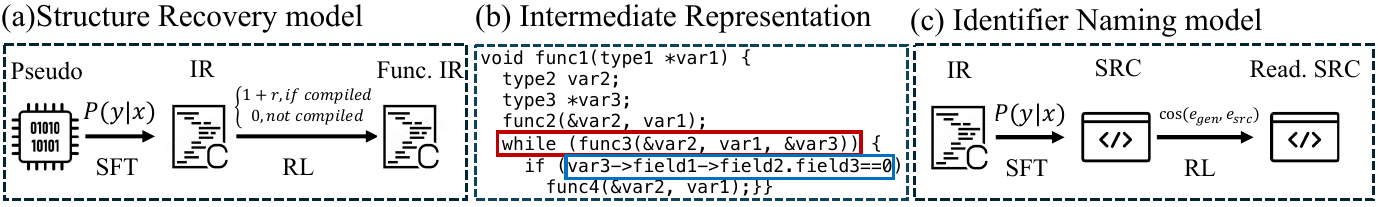}
  \caption{Overview of \method{} with two-phase decompilation process (a) \struc{} and (c) \ident{}. Obfuscated source code (b) serves as the Intermediate Representation (IR) connecting these two phases. For each model, a supervised fine-tuning (SFT) process is performed followed by Reinforcement Learning (RL) with its respective reward.}
  \label{fig:train}
\end{figure*}

Figure~\ref{fig:train} presents the framework of \method{} (\textbf{Sk}eleton-to-\textbf{Sk}in \textbf{Decompile}) which includes a two-phase decompilation process, i.e., 
\struc{} and \ident{} (Section~\ref{sec:decom}) which are realized upon the design of the Intermediate Representation (IR, Sections~\ref{sec:ir} and ~\ref{sec:irgen}), with their respective reward functions to advance the correctness and readability of the final decompiled code (Section~\ref{sec:rl}).

\subsection{Two-phase decompilation process}\label{sec:decom}

We analogize the two-phase decompilation process of \method{} to the human body. In particular, \struc{} refers to  constructing the global code structure, such as loops, conditionals, and data structures, as deriving the program's ``skeleton''. 
\ident{} refers to inferring meaningful names for functions, types, fields, and variables to further reflect actual program semantics.

We formalize \method{} from a probabilistic perspective. In particular, the goal of decompilation is to find the most probable source code ($\rs$) given a low-level representation, i.e., the pseudocode ($\ru$) in our paper. Correspondingly, the decompilation goal can be modeled as maximizing the conditional probability $P(\rs | \ru)$. 
Our core hypothesis is that introducing an intermediate representation ($\ri$) can simplify this task. Using the chain rule of probability, we can decompose the probability
$P(\rs | \ru)$ as $\sum_{\ri} P(\rs | \ri, \ru) \cdot P(\ri | \ru)$.
This decomposition effectively splits the decompilation task into two more manageable sub-tasks, which we illustrate with the example in Figure~\ref{fig:train}(b) and its corresponding pseudocode and source code presented in Figure~\ref{fig:case}(a) and Figure~\ref{fig:case}(c).

\textbf{\struc{}.}
This phase corresponding to $P(\ri | \ru)$ focuses exclusively on translating the syntax and control flow of the low-level pseudocode (Figure~\ref{fig:case}(a)) to a well-formed, high-level IR (Figure~\ref{fig:train}(b)). For example, this task includes identifying that a \verb|while(1)| combined with a \verb|goto LABEL_2| in the pseudocode corresponds to a single, conditional \verb|while()| loop structure in the IR.
This phase also transforms opaque pointer arithmetic, like \verb|*(uint8_t *)(v5 + 1)|, to a clean, nested structure access \verb|var3->field2->field3|.

\textbf{\ident{}:}
This phase corresponding to $P(\rs | \ri, \ru)$ takes the recovered IR (Figure~\ref{fig:train}(b)) and infers meaningful names for variables and functions to produce the final, human-readable source code (Figure~\ref{fig:case}(c)), e.g., transforming a generic call \verb|var3->field2->field3| to the one with more meaningful, semantic names \verb|entry->key->obj.markers|.

A key insight is that once the clean structured IR is recovered, the original, messy pseudocode provides almost no additional information for the naming task. For instance, after recovering the structure \verb|var3->field2->field3|, the model no longer needs the pointer expression \verb|*(uint8_t *)(v5 + 1)| to infer the correct variable names. This insight allows us to make a Markov assumption~\citep{prob_markov}, which simplifies the naming probability from $P(\rs | \ri, \ru)$ to $P(\rs | \ri)$. This simplification yields our final probabilistic model:

\begin{equation}
P(\rs | \ru) \approx \sum_{\ri} P(\rs | \ri) \cdot P(\ri | \ru)
\end{equation}

By decomposing the problem, we create a focused, two-phase process. First, we solve the complex \struc{} challenge ($P(\ri | \ru)$), and then perform the \ident{} task on a clean, abstract representation ($P(\rs | \ri)$). In this way, we reduce the overall complexity for more robust learning and higher-quality decompilation.

\subsection{Intermediate Representation}\label{sec:ir}
The two-phase decompilation process necessitates an intermediate representation (IR) that serves as a bridge between pseudocode and source code. However, designing the IR presents a fundamental challenge, i.e., it must be simple enough to be reliably recovered from pseudocode, yet rich informative enough to enable accurate source code reconstruction.

This challenge naturally frames our problem as an Information Bottleneck (IB) optimization task~\citep{infobot1,infobot2}. In particular, for any information flow pseudocode → IR → source, the intermediate representation acts as a bottleneck that must balance two competing factors, i.e.,  
\textit{compression} and \textit{relevance}. More specifically, compression means that the IR should discard irrelevant details from the pseudocode to make the \struc{} phase tractable. Moreover, relevance refers to that the IR must preserve sufficient information to reconstruct the source code in the \ident{} phase. Ideally, the IR should be maximally inferable from the pseudocode and structurally close to the target source code.
Accordingly, the Information Bottleneck (IB) principle formalizes this trade-off through the objective:
\begin{equation}
\min_{P(i|u)} \; \mathcal{L}_{\mathrm{IB}} = I(\ru;\ri) - \beta I(\ri;\rs),
\label{eq:ib}
\end{equation}

where $I(\ru;\ri)$ measures the mutual information between pseudocode and IR (to be minimized) and $I(\ri;\rs)$ measures the mutual information between IR and source code (to be maximized). Formula~\ref{eq:ib} guides our choice of IR. We thus propose using obfuscated source code~\citep{dobf-cite}, particularly the original source with all identifiers replaced by generic placeholders. Such a representation emerges naturally from the IB objectives. In particular, for the compression objective, the model should distill high-level structural abstractions from the noisy, low-level patterns of the input pseudocode. This process inherently discards irrelevant input details, thus minimizing the mutual information between the input and our IR. Meanwhile, for the relevance objective, the obfuscated code is an ideal structural representation as it can be theoretically recoverable from compiled binary code even when the semantics embodied in original identifiers is lost during compilation. Consequently, this IR preserves the maximum possible relevant information about the source code, thereby maximizing the mutual information between the IR and the source.

Note that the obfuscated code can be automatically generated from source code through identifier obfuscation (Section~\ref{sec:irgen}), making it practical in real world.


\begin{algorithm}[ht]
\caption{Generation of Intermediate Representation (IR)}
\begin{algorithmic}[1]
\REQUIRE Source code $C$, corresponding pseudocode $P$
\ENSURE Obfuscated source code ($IR$)

\STATE Analyze $P$ to extract names that need to be preserved: $F_P$
\STATE Parse $C$ into an abstract syntax tree (AST): $T$
\STATE Initialize rename maps $R[\cdot] \gets \emptyset$ for \texttt{func}, \texttt{type}, \texttt{field}, \texttt{var}
\STATE Initialize counters $cnt[\cdot] \gets 1$ for each identifier type
\STATE Initialize replacement list $\mathcal{L} \gets \emptyset$

\STATE \textbf{Traverse($node$):}
\STATE \quad $(id\_type, name) \gets$ classify $node$
\STATE \quad \textbf{if} $name \notin F_P$ \textbf{then}
\STATE \quad \quad \textbf{if} $name \notin R[id\_type]$ \textbf{then}
\STATE \quad \quad \quad $new \gets id\_type \,\|\, cnt[id\_type]$
\STATE \quad \quad \quad $R[id\_type][name] \gets new$
\STATE \quad \quad \quad $cnt[id\_type] \gets cnt[id\_type] + 1$
\STATE \quad \quad \textbf{end if}
\STATE \quad \quad Append replacement $(start(node), end(node), R[id\_type][name])$ to $\mathcal{L}$
\STATE \quad \textbf{end if}
\STATE \quad \textbf{for each} child $c$ of $node$ \textbf{do}
\STATE \quad \quad Traverse($c$)
\STATE \quad \textbf{end for}

\STATE \textbf{Obfuscate($C, \mathcal{L}$):}
\STATE \quad Sort $\mathcal{L}$ by start position in descending order
\STATE \quad Let $IR$ be a mutable copy of $C$
\STATE \quad \textbf{for each} $(s, e, new)$ in $\mathcal{L}$ \textbf{do}
\STATE \quad \quad Replace substring $IR[s:e]$ with $new$
\STATE \quad \textbf{end for}
\STATE \quad \textbf{return} $IR$

\RETURN $IR$

\end{algorithmic}
\end{algorithm}

\subsection{IR Generation}\label{sec:irgen}
Algorithm 1 illustrates the process of generating the obfuscated code (IR) from the source code. Specifically, the pseudocode is first analyzed to extract all function and type names that should remain unchanged in the obfuscated code, e.g., the standard type \texttt{int} and library function \texttt{memcpy}, which are stored in the reserved list $F_P$  (line 1). The source code is then parsed into an abstract syntax tree (AST) to provide precise identifier positions (line 2). For each identifier category, we initialize renaming maps and counters, as well as an empty replacement list (lines 3–5). We then invoke the recursive procedure \textsc{Traverse} on the root of the AST (lines 6-18). During traversal, each node is classified to determine its identifier type and name (line 7). If the name does not appear in the reserved list $F_P$, a new obfuscated name is generated and stored in the renaming map (lines 8–13). A replacement entry containing the start and end offsets together with the new name is then appended to the replacement list (line 14). The procedure continues recursively on all children of the current node (lines 16-18). After traversal, the replacements are applied (lines 19-25) in \textsc{Obfuscate} where the list is sorted in descending order of start position (line 20) so that later modifications do not shift earlier offsets, and all substitutions are performed on the original code (lines 21–24). Finally, the obfuscated code, namely, IR, is returned (line 26).

\subsection{Enhancement with Reinforcement Learning}\label{sec:rl}
To recover the structured IR from pseudocode and the identifier names from the structured IR, we adopt the sequence-to-sequence (S2S) paradigm, which is adopted in many neural machine translation models that aim to predict the output given the input sequence~\citep{transformers}. This paradigm typically minimizes the cross-entropy (CE) loss for the predicted tokens: $y_i$:$\mathcal{L}_{\text{CE}}(\theta) = - \sum_{i=1}^{N} \log P_{\theta}\!\left( y_i \mid y_{<i}, x \right)$, i.e., calculating the total loss by summing the negative log probabilities of the model correctly predicting each token in a sequence, given all the preceding tokens.

Such CE loss refers to an aggregation of local, token-level prediction errors, serving as a baseline to train an LLM-based decompiler. However, it lacks syntactic and semantic awareness, including assigning equal penalties for unequal errors. For example, a misplaced semicolon, which breaks compilation, might receive a similar penalty to choosing a semantically equivalent but different variable name. 
Therefore, only adopting a supervised model for \struc{} might generate syntactically plausible code that does not compile, limiting the effectiveness of \method{}.

To further enhance \struc{}, after the S2S training, we perform reinforcement learning (RL) to align the outputs with compiler's preference and type constraints such that the generated IR could better represent a compilable and functionally sound program.
Specifically, we design the reward made up of two components. First, for each generated IR, we provide the compiler with the header of the ground-truth IR in order to verify its compilability and grant a reward only upon success, for advancing functional correctness. Additionally, we reward the accurate recovery of placeholder identifiers by computing the Jaccard similarity coefficient between the generated ($I_{\text{gen}}$) and ground-truth sets ($I_{\text{IR}}$). The placeholder recovery reward encourages the model to accurately reconstruct the program's data layout. Formally:

\begin{equation}
r_{\text{placeholder}} = \frac{|I_{\text{gen}} \cap I_{\text{IR}}|}{|I_{\text{gen}} \cup I_{\text{IR}}|},
\qquad
r_{\text{structure}} =
\begin{cases}
0.0, & \text{if } IR \text{ cannot be compiled} \\
1.0 + r_{\text{placeholder}}, & \text{if } IR \text{ can be compiled}
\end{cases}
\end{equation}
Note that using compiler feedback as a reward is feasible and natural. One possible alternative is to build the reward based on the executing unit tests. However, creating unit tests and replicating execution environments for real-world programs is often prohibitively complex and costly.

Similarly, for \ident{}, the CE loss undesirably penalizes cases where identifiers differ superficially but are semantically equivalent, while such differences are negligible from a human’s perspective. To mitigate this, we also perform RL and formulate the corresponding reward as the semantic similarity between the embedded generated code ($\mathbf{e}_{\text{gen}}$) and the reference source code ($\mathbf{e}_{\text{src}}$), measured by the cosine similarity:
\begin{equation}
r_{\text{identifier}} = \cos\!\left(\mathbf{e}_{\text{gen}}, \mathbf{e}_{\text{src}}\right) 
= \frac{\mathbf{e}_{\text{gen}} \cdot \mathbf{e}_{\text{src}}}{\|\mathbf{e}_{\text{gen}}\| \, \|\mathbf{e}_{\text{src}}\|}
\end{equation}

By optimizing for this similarity metric, we encourage the model to generate names that are more semantically aligned with the ground truth, in contrast to the CE loss, which strictly enforces an exact lexical match.

\section{Experiments}

\paragraph{Training Data.}\label{sec:train}
We collected our training corpus from the C programs of Exebench~\citep{exebench} and Decompile-Bench~\citep{decompilebench} datasets. We compiled the source files into binaries for the x86 Linux platform using GCC and Clang~\citep{clang}, applying optimization levels -O0 through -O3. To ensure data quality, we normalized the code by removing all comments and applying clang-format to the source code, while formatting the pseudocode to adhere to the R2I standard~\citep{r2i}. We further employed MinHash-LSH to identify and remove near-duplicates~\citep{minhash}. Following previous reverse engineering practices~\citep{dire,dirty,resym}, we stripped all binaries and used IDA Pro~\citep{idapro} to generate pseudocode (please refer to Appendix~\ref{appen:strip} for stripping examples). This process yielded a comprehensive dataset of approximately 5 million samples, totaling around 2B tokens of pseudocode, 1.5B tokens of IR, and 1.5B tokens of source code. 

\paragraph{Evaluation Benchmarks and Metrics.}
For evaluation, we adopted a set of standard benchmarks widely used in previous studies: HumanEval~\citep{humaneval}, ExeBench~\citep{exebench}, MBPP~\citep{mbpp}, and Github2025~\citep{decompilebench}. These benchmarks were processed using the same compilation pipeline as our training data. To assess the quality of the generated decompiled code, we used three primary metrics, i.e.,  
\textit{R2I}~\citep{r2i}, \textit{GPT-Judge}~\citep{gpt-eval-decompile,gpt-eval-nl}, and \textit{re-executability rate}~\citep{slade,LLM4Decompile}. In particular, R2I measures the relative readability of code structure.
GPT-Judge uses GPT-5-mini~\citep{gpt5} to evaluate the \ident{} effectiveness of the output, with 1 for poor performance to 5 for excellent performance.
For benchmarks that support execution (HumanEval and MBPP), we also measure the re-executability rate~\citep{slade,LLM4Decompile}, which checks if the decompiled code can be successfully re-compiled and passes the original test cases. For tests on stripped binaries, we restore the original function name in the generated code. Note that Exebench is excluded from the evaluation on re-executability rate because the stripping process disrupts its required execution environment. Detailed definitions for each metric and the prompt of GPT-Judge are provided in Appendix~\ref{appen:metric}.

\paragraph{Baselines.}
We compare against GPT-5-mini~\citep{gpt5}, a state-of-the-art commercial model, as well as two leading open-source decompilation models LLM4Decompile~\citep{LLM4Decompile} and Idioms~\citep{idioms}. Other LLM-based decompilers, such as Nova~\citep{nova}, Ref-Decomp~\citep{ref-decompile}, and D-Lift~\citep{dlift}, were not included in our comparison because they do not provide details about their data preprocessing approaches or do not release their models, hindering fair and reproducible evaluations.

\paragraph{Configurations.}
Both the \struc{} and \ident{} models were initialized from the LLM4Decompile-6.7B checkpoint~\citep{LLM4Decompile}. We performed supervised fine-tuning for one epoch using the LLaMA-Factory library~\citep{zheng2024llamafactory} with a batch size of 128 and a learning rate of $3e-6$. For the Reinforcement Learning (RL) phase, we leveraged the GRPO~\citep{deepseekr1} algorithm in the veRL library~\citep{verl} and trained on a random subset of 50,000 samples due to computational constraints. The RL reward for code compilability is verified using Psyche-C~\citep{psychec} to generate headers, and the reward for semantic similarity is measured using qwen-embedding-0.6B~\citep{qwen3embedding}.
All experiments were conducted on clusters of NVIDIA H800-80GB GPUs.
During inference, we used the vLLM~\citep{kwon2023efficient} library for accelerated generation and employed greedy decoding to 
minimize randomness.

\subsection{Main Results}

\begin{table*}[ht]
\centering
\caption{Re-executability results between the studied decompilers.
}
\begin{adjustbox}{width=1\columnwidth}

\begin{tabular}{ccccccccccc}
\toprule
\multirow{2}{*}{\textbf{Re-Executability Rates}}      & \multicolumn{5}{c}{HumanEval}                         & \multicolumn{5}{c}{MBPP}                                                            \\
                       \cmidrule(lr){2-6}                                                                        \cmidrule(lr){7-11}
                       & O0              & O1              & O2              & O3              & AVG             & O0              & O1              & O2              & O3              & AVG             \\
\midrule
GPT-5-mini & 67.07         & 60.78        & 49.63         & 49.56        & 56.75      & 55.70        & 49.33       & 44.13       & 39.74        & 47.23         \\
LLM4Decompile    & 67.07          & 37.25     &  33.58        & 28.32         & 41.71       & 61.56          & 42.42        & 36.90          &     31.32         & 43.05         \\
Idioms                 & 70.73          & 25.49          & 12.41        & 10.62         & 29.81         & 54.78          & 21.58         & 11.60        & 8.06        & 24.01         \\
\method{} & \textbf{86.59}          & \textbf{70.59}         & \textbf{61.31}         & \textbf{57.52}        & \textbf{69.00}         & \textbf{69.76}        & \textbf{62.33}       & \textbf{54.83}       & \textbf{51.58}         & \textbf{59.63}         \\
\bottomrule
\end{tabular}
\end{adjustbox}

\label{table:exe_main}
\end{table*}
\begin{table*}[ht]
\centering
\caption{R2I results between the studied decompilers with the compilation optimization levels O0, O3 and the averaged results on -O\{0,1,2,3\}.}
\begin{adjustbox}{width=\textwidth}

\begin{tabular}{cccccccccccccccc}
\toprule
\multirow{2}{*}{\textbf{R2I }} & \multicolumn{3}{c}{HumanEval} & \multicolumn{3}{c}{MBPP} & \multicolumn{3}{c}{ExeBench} & \multicolumn{3}{c}{GitHub2025} \\
\cmidrule(lr){2-4} \cmidrule(lr){5-7} \cmidrule(lr){8-10} \cmidrule(lr){11-13}
& O0 & O3 & AVG & O0 & O3 & AVG & O0 & O3 & AVG & O0 & O3 & AVG \\
\midrule
GPT-5-mini & 50.52 & 40.03 & 42.33  & 52.13 & 41.66 & 44.14 & 46.91 & 40.63 & 45.25 & 34.82 & 26.44 & 28.95 \\
LLM4Decompile & 64.23 & 68.02 & 68.71  & 66.10 & 68.66 &69.19& 62.64 & 52.79 & 57.02 & 49.65 & 49.78 & 47.95\\
Idioms & \textbf{68.50} & 45.43 & 53.10 & \textbf{72.11} & 45.55 & 55.73 & 66.77 & 52.17 & 60.40 & 63.50 & 58.21 & 57.97\\
\method{} & 64.92 & \textbf{72.70} & \textbf{70.27}  & 66.23 & \textbf{74.43} & \textbf{71.66} & \textbf{69.96} & \textbf{73.55} & \textbf{71.50} & \textbf{76.46} & \textbf{74.51} & \textbf{74.99} \\
\bottomrule
\end{tabular}

\end{adjustbox}

\label{table:r2i_main}
\end{table*}
\begin{table*}[ht]
\centering
\caption{GPT-Judge results between the studied decompilers.}
\begin{adjustbox}{width=\textwidth}

\begin{tabular}{cccccccccccccccc}
\toprule
\multirow{2}{*}{\textbf{GPT-Judge }} & \multicolumn{3}{c}{HumanEval} & \multicolumn{3}{c}{MBPP} & \multicolumn{3}{c}{ExeBench} & \multicolumn{3}{c}{GitHub2025} \\
\cmidrule(lr){2-4} \cmidrule(lr){5-7} \cmidrule(lr){8-10} \cmidrule(lr){11-13}
& O0 & O3 & AVG & O0 & O3 & AVG & O0 & O3 & AVG & O0 & O3 & AVG \\
\midrule
GPT-5-mini & 4.49 & \textbf{4.07} &4.23  & \textbf{4.35} & 3.88 & 4.08 & \textbf{2.53} & 2.33 & 2.37& 3.04 & 2.86& 2.87 \\
LLM4Decompile & 3.88 & 3.29& 3.42  & 3.81 & 3.22 & 3.41 & 2.47 & 2.12 & 2.22 & 2.52& 2.56 & 2.62\\
Idioms & 4.30 & 2.70 & 3.22 & 4.07 & 2.61 & 3.13 & 2.46 & 1.71 & 2.01 & 2.51 & 2.10 & 2.18\\
\method{} & \textbf{4.51} & 4.05& \textbf{4.24}  & 4.31 & \textbf{3.95} & \textbf{4.12} & 2.48 & \textbf{2.47} & \textbf{2.42} &\textbf{3.05} & \textbf{3.02} & \textbf{3.06} \\
\bottomrule
\end{tabular}

\end{adjustbox}

\label{table:gpt_main}
\end{table*}

Table~\ref{table:exe_main} compares the re-executability rates of the studied decompilers on the HumanEval and MBPP datasets across different optimization levels (O0-O3). Notably, \method{} achieves the highest performance, surpassing the best-performing baseline GPT-5-mini by 21.6\% and 26.3\% averagely on each dataset. Specifically, to the best of our knowledge, this is the first model that preserves the functionality of binaries and reaches an average re-executability of $\sim$70\% and $\sim$60\% of HumanEval and MBPP cases, underscoring the advantage of decomposing decompilation into two sub-tasks. 

Table~\ref{table:r2i_main} presents the R2I results of the studied decompilers. \method{} consistently outperforms all the baselines. The improvements are particularly significant in the recovery of program structures from real-world binaries. Specifically, on the ExeBench and GitHub2025 datasets, \method{} achieves performance gains of 18.4\% and 29.4\% over the best-performing baseline Idioms.

The effectiveness of \ident{}, as evaluated by GPT-Judge, is presented in Table~\ref{table:gpt_main}, where \method{} produces high-quality names on both the HumanEval and MBPP datasets, achieving scores of 4.24 and 4.12 out of 5, respectively. Furthermore, when applied to the real-world datasets, \method{} demonstrates an advantage over the existing techniques, outperforming GPT-5-mini by 2.1\% and 6.7\%.

\subsection{Ablations}
\begin{table*}[ht]
\centering
\caption{Re-executability results between the \method{} variants.
}
\begin{adjustbox}{width=1\columnwidth}

\begin{tabular}{ccccccccccc}
\toprule
\multirow{2}{*}{\textbf{Re-Executability Rates}}      & \multicolumn{5}{c}{HumanEval}                         & \multicolumn{5}{c}{MBPP}                                                            \\
                       \cmidrule(lr){2-6}                                                                        \cmidrule(lr){7-11}
                       & O0              & O1              & O2              & O3              & AVG             & O0              & O1              & O2              & O3              & AVG             \\
\midrule
pseudo-src & 73.78          & 54.25        & 48.91         & 42.48        & 54.86       & 58.37        & 48.97        & 42.77       & 39.93         & 47.51         \\
pseudo-ir    & 78.66         & 65.35      & 54.01        & 52.21         & 62.56        & 55.29          & 49.33         & 43.22          &     41.02          & 47.25        \\
pseudo-ir-rl    & \textbf{87.80}         & 69.28     & 59.85         & \textbf{58.41}         & 68.84         & 68.35          & 59.88         & 52.25          &     47.62          & 57.06         \\
pseudo-ir-src                 & 78.66          & 66.01         & 55.47         & 54.86          & 63.75          & 60.95          & 55.03         & 48.34       & 46.89       & 52.83          \\
pseudo-ir-src-rl & 86.59          & \textbf{70.59}         & \textbf{61.31}         & 57.52         & \textbf{69.00}        & \textbf{69.76}       & \textbf{62.45}         & \textbf{54.83}       & \textbf{51.58}        & \textbf{59.63}         \\
\bottomrule
\end{tabular}
\end{adjustbox}

\label{table:exe_abla}
\end{table*}

In the ablation study, we designed a series of \method{} variants to indicate the individual effects of its major components, including the task decomposition and the crafted reward (RL) as follows. 

\begin{itemize}[leftmargin=*, topsep=0pt]

\item{pseudo-src}: This model represents a direct, end-to-end approach to decompile from pseudo to source code. It is trained using SFT with the same training data used in \method{}.
\item{pseudo-ir}: This model is trained with SFT to convert pseudocode to IR for the evaluation on the effectiveness of \struc{}.
\item{pseudo-ir-src}: This model is trained with SFT to convert IR to source code. The output IR from the \struc{} phase serves as its input for evaluating the effectiveness of \ident{}. Note that the training cost of a direct approach, \texttt{pseudo-src} and decomposed approach, \texttt{pseudo-ir} with \texttt{pseudo-ir-src}, are similar to ensure fair comparison.
\item{pseudo-ir-rl}: Based on \texttt{pseudo-ir}, the model is further tuned with RL on compiler feedback.
\item{pseudo-ir-src-rl}: This model is the complete version of \method{} which integrates the decomposed, two-phase framework enhanced with the RL for both phases.
\end{itemize}

Table~\ref{table:exe_abla} presents our ablation study results in terms of the re-executability rates. Noticing that \texttt{pseudo-src} establishes a baseline performance with re-executability rates of 54.86\% and 47.51\% on HumanEval and MBPP dataset, splitting the decompiliation process into \struc{} and \ident{} (\texttt{pseudo-ir-src}) could increase corresponding scores to 63.75\% and 52.83\% respectively. This validates that tackling decompilation as two simpler sub-tasks is indeed a more effective strategy. Notably, even the \struc{} model alone (\texttt{pseudo-ir}) surpasses the \texttt{pseudo-src} baseline on HumanEval, highlighting that an independent phase on program structure recovery is a critical factor.
Moreover, \texttt{pseudo-ir-rl} achieves dramatic performance gains of 10.0\% and 20.8\% over the supervised-only model \texttt{pseudo-ir}, indicating the benefit of crafted rewards.
At last, \textit{pseudo-ir-src-rl} achieves the best performance, demonstrating that each component of \method{} is critical and combining them together is essential for optimizing the performance. We observe similar trends for the R2I and GPT-Judge results and present them in Appendix~\ref{appen:result} due to the page limit.

\subsection{Case Study}
\begin{figure*}[hbt]
  \centering
  \includegraphics[width=1.0\textwidth]{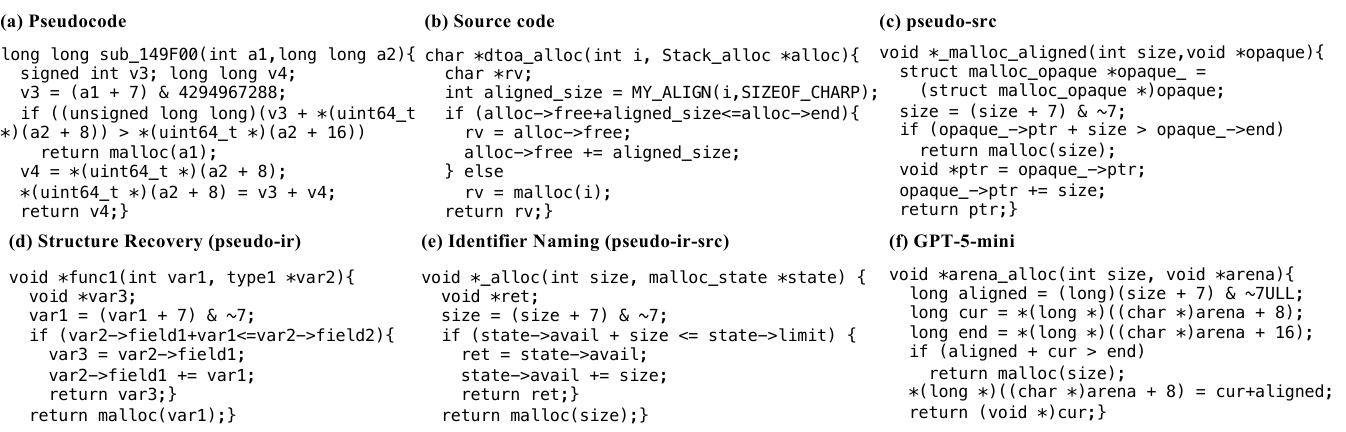}
  \caption{A case study on a memory allocation function with (a) pseudocode, (b) source code, (c) decompilation result from \texttt{pseudo-src} in Table~\ref{table:exe_abla}, (d) \struc{} result from \texttt{pseudo-ir} in Table~\ref{table:exe_abla}, (e) \ident{} result from \texttt{pseudo-ir-src} in Table~\ref{table:exe_abla}, and (f) decompilation from GPT-5-mini.} 
  \label{fig:finalcase}
\end{figure*}
 Figure~\ref{fig:finalcase} presents a case study on a memory allocation function. A direct decompilation in Figure~\ref{fig:finalcase}(c) produces non-intuitive code that relies on an explicit type cast from a generic \verb|void * opaque|. It also incorrectly identifies the \verb|free| field as \verb|ptr|. GPT-5-mini in Figure~\ref{fig:finalcase}(f) fails to reconstruct the data structure, and represents data access with a low-level pointer offset \verb|(char *)arena + 8| instead. In contrast, the \struc{} phase of \method{} in Figure~\ref{fig:finalcase}(d) successfully recovers the essential control flow, conditions, and data structures. Building on the clean recovered structure from Figure~\ref{fig:finalcase}(d), the \ident{} phase in Figure~\ref{fig:finalcase}(e) further enhances readability by assigning meaningful names to identifiers, such as inferring \verb|available| and \verb|state| as semantically appropriate names for the original \verb|free| and \verb|alloc|, leading to a structurally accurate and semantically rich result.

\section{Conclusion}
In this work, we propose \method{} which decomposes the binary decompilation task into two phases. First, it recovers the program's ``skeleton'', i.e., its functional structure, using an Intermediate Representation and compiler-guided Reinforcement Learning. Second, it recovers the program's ``skin'', i.e., naming identifiers, with a separate reward on semantic similarity to improve readability. 
Experimental results show that \method{} is the first to achieve the average re-executability rate of approximately 70\% on HumanEval and 60\% on MBPP datasets. It also achieves a 21.6\% average re-executability rate gain over GPT-5-mini on HumanEval and 29.4\% average R2I improvement over Idioms on the GitHub2025 benchmark. In conclusion, \method{} significantly outperforms the existing techniques in producing functionally correct and human-readable decompilation code.

\section*{Reproducibility statement}
To ensure full reproducibility of our results, we have made all associated artifacts publicly available in an anonymous GitHub repository. This repository contains the complete source code for our model implementation, training scripts, and evaluation protocols. We also provide the processed testing data, along with scripts for data preparation. For ease of use, pre-trained model weights are also released. The README.md file in the repository offers a step-by-step guide to set up the environment, and replicate the key results presented in this paper.

\section*{Ethics}
\method{} was developed under strict ethical guidelines. It is intended for use in legitimate scenarios, such as academic research, debugging, and recovering a company's own lost source code, where permission is granted or copyright does not apply. To support this, the model was trained exclusively on open-source code from public benchmarks and permissively licensed repositories, e.g., MIT, BSD, and Apache 2.0~\citep{stackv2}. Notably, commercial software remains well-protected by obfuscation methods that make effective decompilation infeasible~\citep{LLM4Decompile}, thus limiting the potential for misuse.



\bibliography{iclr2026_conference}
\bibliographystyle{iclr2026_conference}

\appendix
\section{Appendix}
\subsection{Strip}\label{appen:strip}
\begin{figure*}[hbt]
  \centering
  \includegraphics[width=\textwidth]{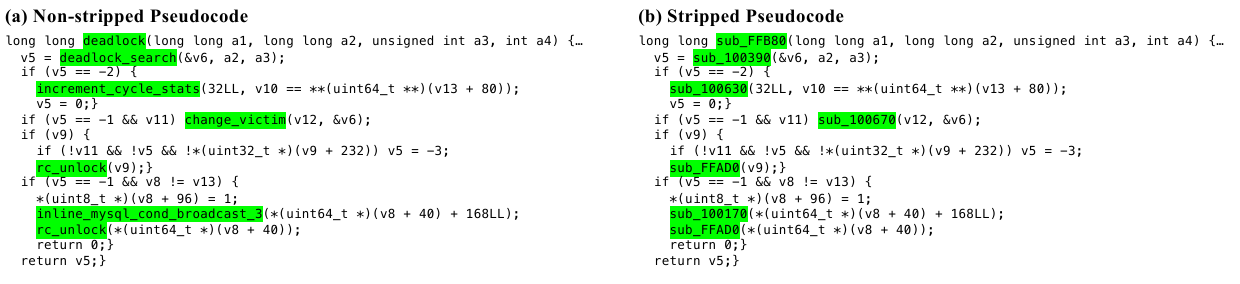}
  \caption{An example with its (a) not striped pseudocode, (b) striped pseudocode}
  \label{fig:strip}
\end{figure*}

Stripping is the process of removing non-essential information from binary executable files and object files~\citep{name-recover1, name-recover2, hext5, decompilation4_gnn}. This information, primarily intended for debugging and analysis, is not required for the program's actual execution. The data typically removed includes 
\textit{Symbol Tables} and \textit{Debugging Information}. Specifically, symbol tables contain the names and addresses of functions, global variables, and other objects within the program. Debugging information refers to the extra data generated by the compiler (e.g., with the -g flag in GCC) that maps the compiled machine code back to the original source code lines, variable names, and data structures.

Stripping binaries is a \textbf{common and standard practice}~\cite{dire,dirty,resym,gpt-eval-decompile}, particularly for software deployed to production environments, as it ensures size reduction and enhances security. The removal of symbols and debugging information can significantly decrease the size of an executable file. A stripped binary is considerably more difficult for reverse engineers. Without meaningful function and variable names, an attacker or competitor must invest significantly more time and effort to understand the program's internal workings, business logic, or potential vulnerabilities. 

The pseudocode snippets in Figure~\ref{fig:strip} offer an illustration of stripping on a program. In particular, the non-stripped pseudocode in Figure~\ref{fig:strip}(a) is significantly more readable to a human analyst. It features descriptive function names such as \verb|deadlock|, \verb|deadlock_search|, \verb|increment_cycle_stats|, \verb|change_victim|, and \verb|rc_unlock|. These names provide immediate insight into the potential purpose of the code, suggesting it is part of a system designed to detect and handle deadlocks in a database context, possibly related to MySQL as indicated by \verb|inline_mysql_cond_broadcast_3|. On the other hand, the stripped pseudocode in Figure~\ref{fig:strip}(b) is obfuscated. The meaningful function names have been replaced with generic, tool-generated placeholders like \verb|sub_FFB80|, \verb|sub_100390|, \verb|sub_100630|, and \verb|sub_FFAD0|. These names are derived from the memory addresses of the functions and offer no clues about their functionality. An analyst examining this code would face a much steeper challenge in deciphering the program's logic and intent.

\begin{figure*}[hbt]
  \centering
  \includegraphics[width=\textwidth]{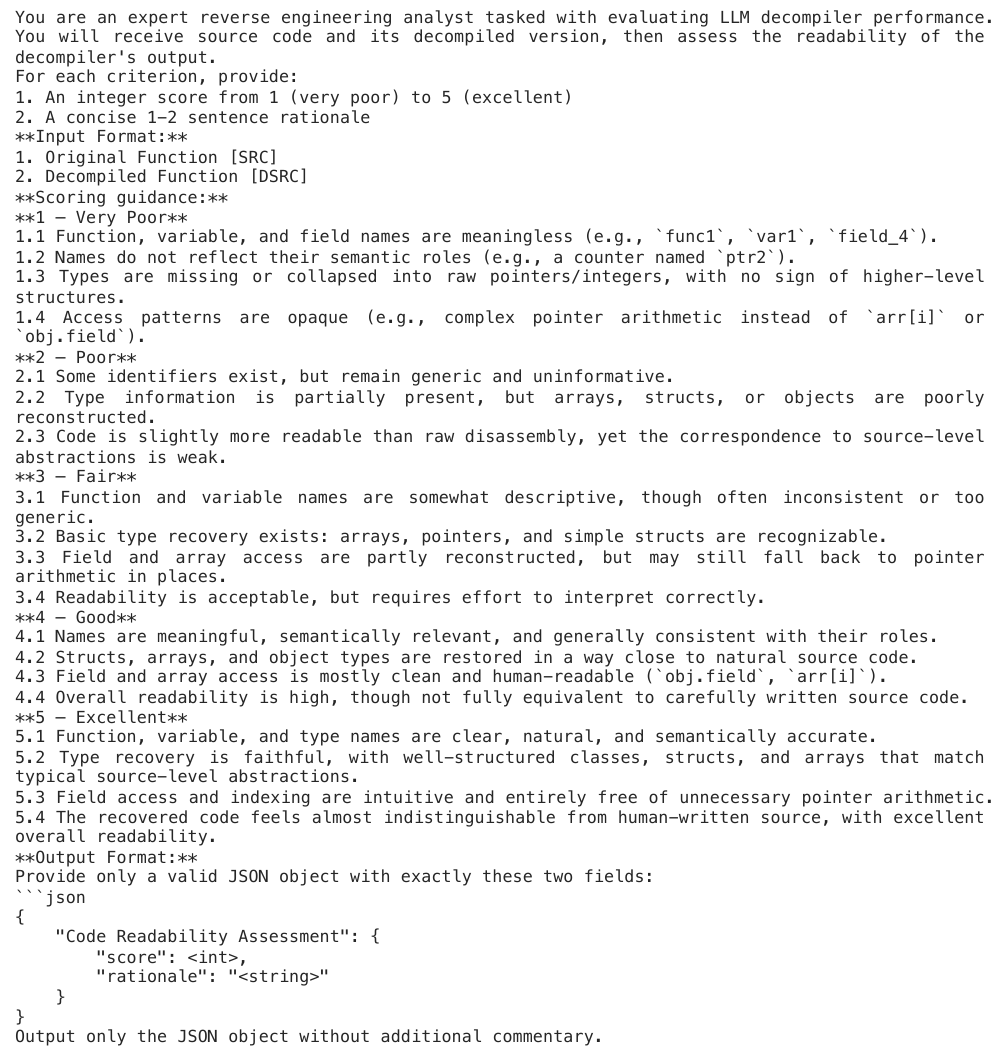}
  \caption{GPT-Judge prompt for a qualitative assessment of \ident{} effectiveness.}
  \label{fig:prom}
\end{figure*}

\subsection{Metrics}\label{appen:metric}
The Relative Readability Index (R2I)~\citep{r2i} is a quantitative metric for evaluating and comparing the readability of decompiled C code, producing a normalized score between 0 and 1. It functions by constructing an Abstract Syntax Tree (AST) for each output, extracting predefined features, and calculating a weighted score.However, the original R2I implementation introduces a significant bias. It discards an entire data sample if any single decompiler's output fails to be parsed by the \texttt{pycparser}~\citep{pycparser} library. This is problematic because \texttt{pycparser} often fails on code containing user-defined types and functions, skewing the evaluation towards simpler programs. To create a more robust and unbiased metric, we modified the process. First, we use \texttt{pschec}~\citep{psychec} to generate headers, improving the likelihood of successful parsing. More importantly, if a specific output still fails to parse, we assign it a score of 0 instead of discarding the entire sample. This allows us to evaluate the other parsable outputs for that program, ensuring a more comprehensive and fair assessment.

Re-executability is a widely adopted metric in decompilation that evaluates the functional equivalence between an original source function and its decompiled output~\citep{LLM4Decompile,slade,self-decompile,nova,ref-decompile}. Ideally, this means the decompiled function should produce the same output as the original function for every conceivable input. However, since testing every input is impossible, we use a practical approach. We run a set of predefined unit tests on both the original code and the decompiled code. If the outputs match for every single test case, we consider the decompilation successful and ``re-executable''. This same concept is often called I/O accuracy or pass rate~\citep{slade,nova}.

We leverage GPT-Judge to assess the \ident{} effectiveness for the decompilers. GPT-Judge has become increasingly adopted for evaluating LLM-based decompilers~\citep{LLM4Decompile,decompilebench,decompilebench-qing}. In particular, we use GPT-5-mini~\citep{gpt5} as an automated evaluator, which is prompted to perform a comparative analysis of the decompiled output and the original source code, specifically focusing on the quality of the recovered identifiers. It provides a rating on a 5-point scale, with 1 for poor performance to 5 for excellent performance. The exact prompt used in our evaluation is detailed in Figure~\ref{fig:prom}.

\subsection{Addtional Results}\label{appen:result}
\begin{table*}[ht]
\centering
\caption{R2I results between the \method{} variants. Note that since R2I evaluates decompiled code in a relative context quantitatively~\citep{r2i}, its values can vary significantly for the same decompiler when compared with different baselines.}
\begin{adjustbox}{width=\textwidth} 

\begin{tabular}{cccccccccccccccc}
\toprule
\multirow{2}{*}{\textbf{R2I }} & \multicolumn{3}{c}{HumanEval} & \multicolumn{3}{c}{MBPP} & \multicolumn{3}{c}{ExeBench} & \multicolumn{3}{c}{GitHub2025} \\
\cmidrule(lr){2-4} \cmidrule(lr){5-7} \cmidrule(lr){8-10} \cmidrule(lr){11-13}
& O0 & O3 & AVG & O0 & O3 & AVG & O0 & O3 & AVG & O0 & O3 & AVG \\
\midrule
pseudo-src & 54.62 & 57.77 &56.47  & 56.53 & 54.43 & 55.83 & 59.11 & 49.34 & 55.15& 54.41 & 51.71 & 53.17 \\
pseudo-ir & 51.85 & 60.25 & 56.39  & 54.81 & 55.73 & 55.26 & 55.86 & 53.96 & 55.18 & 58.31 & 55.30 & 56.46\\
pseudo-ir-rl & 55.51 & 60.76 & 57.53 & 54.49 & 58.39 & 57.36 & 59.68 & 60.82 & 60.92 & 59.24 & 56.95 & 57.15\\
psuedo-ir-src & 53.58 & 59.52 & 57.10  & 55.22 & 56.30 & 55.80 & 56.04 & 55.50 & 55.73 & 59.36 & 56.06 & 57.33 \\
psuedo-ir-src-rl & 56.41 & 59.01 & 57.49  & 54.68 & 59.28 & 57.75 & 59.50 & 60.59 & 61.06 & 59.70 &57.56 & 57.73 \\
\bottomrule
\end{tabular}

\end{adjustbox}

\label{table:r2i_abla}
\end{table*}
\begin{table*}[ht]
\centering
\caption{GPT-Judge results between the \method{} variants}
\begin{adjustbox}{width=\textwidth}

\begin{tabular}{cccccccccccccccc}
\toprule
\multirow{2}{*}{\textbf{GPT-judge}} & \multicolumn{3}{c}{HumanEval} & \multicolumn{3}{c}{MBPP} & \multicolumn{3}{c}{ExeBench} & \multicolumn{3}{c}{GitHub2025} \\
\cmidrule(lr){2-4} \cmidrule(lr){5-7} \cmidrule(lr){8-10} \cmidrule(lr){11-13}
& O0 & O3 & AVG & O0 & O3 & AVG & O0 & O3 & AVG & O0 & O3 & AVG \\
\midrule
pseudo-src & 4.45 & 3.80 &4.05  & 4.23 & 3.89 & 4.03 & 2.66 & 2.30 & 2.37& 3.08 &2.89 & 3.00 \\
pseudo-ir & 2.88& 2.69 & 2.74  & 2.78 & 2.64 & 2.72 & 1.96 & 1.73& 1.75 & 2.42 & 2.23 & 2.34\\
pseudo-ir-rl & 2.93 & 2.69 & 2.79 & 2.80 & 2.67 & 2.73 & 1.97 & 1.73 & 1.77 & 2.43 & 2.32 & 2.35\\
pseudo-ir-src & 4.48 & 3.99& 4.16  & 4.26 & 3.94 & 4.09 & 2.47 & 2.45 & 2.38 & 3.02 & 2.96 & 3.03 \\
pseudo-ir-src-rl & 4.51 & 4.05& 4.24  & 4.31 & 3.95 & 4.12 & 2.48 & 2.47 & 2.42 &3.05 & 3.02 & 3.06 
\\
\bottomrule
\end{tabular}

\end{adjustbox}

\label{table:gpt_abla}
\end{table*}

We present ablation results for structural readability (R2I) and identifier quality (GPT-Judge) in Table~\ref{table:r2i_abla} and Table~\ref{table:gpt_abla}, respectively. The R2I scores in Table~\ref{table:r2i_abla} exhibit a consistent trend with our re-executability findings (Table~\ref{table:exe_abla}), further indicating that both task decomposition and specialized rewards improve structural recovery. Table~\ref{table:gpt_abla} presents a similar trend for \ident{}. As expected, the \struc{} models score poorly on this metric since they are explicitly designed not to restore original names. Overall, the results altogether validate the effectiveness of our decomposed approach.

\subsection{The Use of Large Language Models}
During the preparation of this manuscript, we utilized a Large Language Model (LLM) only for assistance with writing. The LLM's role was strictly limited to proofreading, correcting grammatical errors, and improving the clarity and readability of the text. The core research ideas, methodologies, and conclusions presented in this paper were conceived and developed entirely by the authors.

\end{document}